\documentclass{elsart}
\newcommand{\eb}{\begin{eqnarray}}
\newcommand{\ee}{\end{eqnarray}}
\usepackage{amssymb}

\begin{document}

\begin{frontmatter}



\title{Ferromagnetic one dimensional Ti atomic chain}


\author{Jisang Hong}

\address{Department of Physics, Pukyong National University,
Busan 608-737, Korea}

\begin{abstract}
Using the full potential linearized augmented plane wave (FLAPW)
method, we have explored the magnetic properties of one dimensional
(1D) Ti atomic chain. Astonishingly, we for the first time observed
that the 1D Ti atomic chain has ferromagnetic ground state even on
NiAl(110) surface although the Ti has no magnetic moment in bulk or
macroscopic state. It was found that the physical property of direct
exchange interaction among Ti atoms occurred in free standing state
is well preserved on NiAl(110) surface and this feature has an
essential role in ferromagnetism of 1D Ti atomic chain. It was shown
that the m=$|2|$ state has the largest contribution to the magnetic
moment of Ti atom grown on NiAl(110) surface. In addition, we found
that the magnetic dipole interaction is a key factor in the study of
magnetic anisotropy, not the magnetocrystalline anisotropy arising
from spin-orbit interaction.
\end{abstract}

\end{frontmatter}

% main text
\section{Introduction}
It is well known that ultra small nano scale materials display many
peculiar phenomena not observed in macroscopic state and the study
of nano scale material has become the most interesting issue. These
days, due to remarkable progress in the field of atomic manipulation
technique one can even build 1D nanosystems. For instance, 1D Co
ferromagnetic (FM) nanowire was grown on Pt(997) step surface and a
magnetic anisotropy energy was measured \cite{pt}. Also, 1D Au
chains through scanning tunneling microscopy tip control were
prepared on NiAl(110) surface and the electronic structures were
investigated \cite{Au}. In addition, a giant magnetic anisotropy was
obtained in single Co atom and nanoparticles \cite{pt1}. Overall,
the ultra small materials display many exciting physical properties
and a great amount of research effort has been focused on these
materials.

Especially, if the spin degree of freedom is included one can expect
more rich physical phenomena. Indeed, one may easily obtain many
interesting experimental data extracted from nano magnetic materials
since the extensive progress has been achieved in the filed of nano
magnetism. However, the fully quantitative theoretical
investigations are still in early stage. In the theoretical view
point for magnetic properties of 1D nanostructures, we already
presented the results of magnetic anisotropy calculations of
infinite Co nano chains and interesting size dependence of magnetic
properties in finite Co nano wires \cite{we1,we2}. It is well
understood that most of physical properties of materials are very
sensitive to the change in underlying electronic structure and
particularly in ultra small size the effect will be very
significant. Therefore, we can anticipate that the ultra small nano
scale materials show unexpected physical properties different from
those observed in bulk or macroscopic states such as half metallic
character in suspended Co nanowires \cite{half}.

In one dimensional geometry the electronic structure will be
substantially altered due to the reduction of dimension and several
theoretical investigations were performed to explore the peculiar
magnetic properties of 1D nanostructures. For instance, it was
theoretically presented that a Pd atom in unsupported one
dimensional nanowire structure has magnetic moment \cite {pd}. On
the other hand, it was shown that three dimensional nanowires of
non-magnetic (NM) simple metals, for instance Na, Cs, and Al, have
ferromagnetic ground state at certain critical radii although the
calculations are based on rather primitive model \cite{jellium}. It
is quite clear that Cu is non-magnetic material in bulk state. But,
very recent quantum conductance measurement indicated that pure 1D
Cu nano wire has spin polarized electronic band structure due to
oxygen effect \cite{cu,cu1} and our theoretical calculations
revealed that the pure 1D Cu atomic chain has no magnetic moment for
various inter-atomic distances. But, very surprisingly it was
demonstrated that the spin polarization can arise in 1D CuO diatomic
nanowire when the inter-atomic distance between Cu and O is larger
than equilibrium distance \cite{cuo}. These theoretical and
experimental results imply that the reduction of dimension may even
affect the magnetic property of material.

So far, most of experimental data found in nano magnetic systems the
sample materials were basically magnetic in bulk or macroscopic
state and simply the size of material was greatly reduced. Neither
experimentally nor theoretically investigated the possibility of
achieving magnetic state in purely 1D atomic scale from non-magnetic
material in macroscopic size. But, as mentioned above the electronic
structure of ultra small material in low dimension is quite
different from the one found in bulk or macroscopic state.
Therefore, it will be of interest to explore whether or not we can
find magnetic moment in reduced dimension through fully quantitative
calculations, whereas the material is non-magnetic in bulk state. In
this spirit, we considered well known non-magnetic element Ti and
investigated the existence of ferromagnetic ground state in one
dimensional atomic chain structure.

\section{Numerical Methods}
The full potential linearized augmented plane (FLAPW) method was
employed in our calculations. Therefore, no shape approximation is
assumed in charge, potential, and wavefunction expansions. We treat
the core electrons fully relativistically. The generalized gradient
approximation was employed to describe exchange correlation
\cite{gga}. Spherical harmonics with $l_{max}=8$ were used to expand
the charge, potential, and wavefunctions in the muffin tin region.
Energy cut offs of 225 Ry and 13.69 Ry were implemented for the
plane wave star function and basis expansions in the interstitial
region. The NiAl(110) substrate was simulated by 7 layers of
NiAl(110) slab. The 1D atomic chain is placed along the $x$ axis. We
impose periodic boundary condition along the lateral direction ($y$
axis) for computational purpose. The lattice constants of  a=5.454
a.u. and b=15.425 a.u. were used for NiAl(110) substrate. The same
lattice parameters were assumed for the calculations of free
standing nanowire. When we check the possibility of
antiferromagnetic ordering in the 1D Ti nanowire, the size of unit
cell is increased twice in the x-direction.

\section{Results and Discussions}
First, we explored the existence of magnetic phase in unsupported Ti
nanowire with the inter-atomic distance of NiAl(110) surface as
shown in above. Very surprisingly, it was obtained that the total
energy of ferromagnetic phase in free standing 1D Ti nanowire is
lower than that of non-magnetic state. The energy gain resulting
from the spin polarization was 539 meV/atom. In addition, the
calculated magnetic moment of Ti atom was 1.88 $\mu_B$. Based on
this promising result, we performed more realistic calculations,
{\it i.e.} the 1D Ti atomic chain grown on NiAl(110) surface. In
Fig. 1, the structure of unit cell considered in our calculations is
schematically illustrated. The physical property of nanoscale
material is sensitive to underlying electronic structure and the
electronic structure is dependent on atomic structure of material.
It is therefore important to achieve well optimized atomic positions
for quantitative calculations. The optimized atomic structure was
found with FLAPW calculations guided by force and energy
minimization procedures. In Table 1, the vertical positions of Ti
and substrate atoms are presented. As one can see, the surface layer
shows buckling state and this behavior in NiAl(110) surface is
experimentally observed \cite{Au}. Nonetheless, one can see that the
subsurface layers are very flat.
\begin{table}[h]
\caption{Vertical positions of Ti and substrate atoms in atomic
unit.}
%\label{1}
\begin{tabular}{lccccccccc} \hline
atom&&Ni&&&Al&&Ti\\ \hline
type&1&2&3&1&2&3& \\
distance&11.598&11.508&7.712&11.708&7.712&7.712&15.504\\ \hline
\end{tabular}
\end{table}
Now, the main issues is to investigate if the 1D Ti nanowire can
maintain ferromagnetic phase in ground state. For this purpose, we
consider the 1D Ti nanowire on NiAl(110 ) for NM, FM, and
antiferromagnetic (AFM) states, respectively and calculate their
total energies. For convenience, the total energy of FM state is
given as zero for reference. The calculated results are presented in
Table II. We found that the energy gain due to spin polarization
from NM to FM state is 245 meV/cell and furthermore it was obtained
that the FM state is more stable than AFM state. Overall, we
conclude that the 1D Ti atomic chain has ferromagnetic ground state
on NiAl(110) surface. The calculated magnetic moment of Ti atom was
1.24 $\mu_B$, whereas no induced magnetic moment was observed in any
other atoms in substrate. To the best of our knowledge neither
experimentally nor theoretically reported the existence of
ferromagnetic ground state in 1D Ti nanowire so far. Consequently,
we believe that our theoretical calculations are the first
prediction that one may observe ferromagnetic phase in 1D Ti atomic
chain grown NiAl(110) surface.
\begin{table}[h] \caption{Total energies for 1D
Ti/NiAl(110) in meV/cell.}
%\label{1}
\begin{tabular}{lccc} \hline
&NM&FM&AFM\\ \hline
Energy&245&0&200\\
\hline
\end{tabular}
\end{table}

Two physical mechanisms can account for the ferromagnetism in 1D Ti
atomic chain; direct exchange interaction among Ti atoms and
indirect one mediated through substrate electrons. It is trivial
that the direct exchange interaction among Ti atoms is the origin of
ferromagnetism for unsupported nanowire. When the Ti chain is grown
on certain substrate, one should take into account the effect of
indirect exchange interaction through substrate materials. Indeed,
the importance of indirect exchange interaction in two dimensional
Mn monolayer on Nb(001) surface was discussed in our theoretical
calculations \cite{mn}. In the present system we observed that the
energy gain due to spin polarization is substantially reduced
compared to that found in free standing structure and this
suppression is resulting from hybridization with substrate atoms,
but there was no meaningful induced magnetic moment in surface Ni
and Al atoms. Nonetheless, the ferromagnetic ground state is still
maintained even on the NiAl(110) surface. We further analyzed the
changes in charge transfer of Ti atom occurred in muffin tin (MT)
region and found no meaningful disparity in the total number of
charges for Ti atom in both NM and FM states (for FM state, the sum
of majority and minority spin electrons in both free standing and on
NiAl(110)). The only modification realized in Ti/NiAl(110) structure
is the charge transfer about  0.3 electrons from majority to
minority spin state and the spin polarization is originated from the
spin splitting in d-state. We therefore believe that the direct
exchange interaction is a key factor in ferromagnetism of 1D Ti
nanowire although the magnitude of exchange interaction is affected
by substrate atoms.

Since it was shown that the 1D Ti nanowire has ferromagnetic ground
state, it is necessary to explore the density of states (DOS)
features. We present the DOS of Ti adatom for free state and on
NiAl(110) surface in Figs. 2(a) and 2(b), respectively. As shown in
Fig. 2(a) for unsupported nanowire, one can see that there are no
available states for minority spin electrons at the Fermi level and
this indicates that the 1D Ti atomic chain is half metallic. In
Ti/NiAl(110) system, we found broadening of DOS due to hybridization
with substrate elements and the half metallic feature is
disappeared. It is clearly displayed that the d holes in majority
states are increased, while the minority spin holes are decreased.
This can nicely account for the suppression of magnetic moment in
Ti/NiAl(110). To provide a comprehensive picture for the peculiar
ferromagnetism of 1D Ti atomic chain, we calculated the m-resolved
DOS. Figs. 3(a) and 3(b) show the m-resolved DOS of Ti for free
standing and on NiAl(110) surface, respectively. The rough
estimation reveals that the number of electrons below the Fermi
level of $|m|$=2, $|m|$=1, and $|m|$=0 states are 0.76, 0.74, and
0.35 for unsupported case, respectively. This implies that the spin
polarizations of $|m|$=1 and $|m|$=2 states play an important role
for the ferromagnetism of free standing Ti nanowire. Substantial
changes in m-resolved DOS of Ti were found on NiAl(110). The half
metallic behavior observed in free standing structure is disappeared
and one can find sizable minority spin states at the Fermi level
instead. This stems from $|m|$=1 and $|m|$=2 orbitals , while the
m=0 state, {\it i.e.} $d_{3z^2-r^2}$ orbital, does not play any
role. The minority spin electrons mainly reside in $|m|$=2 states,
{\it i.e.} $d_{x^2-y^2}$ and $d_{xy}$. Our calculations show that
the number of minority spin electrons below the Fermi level in
$|m|$=2 state are about 0.1, whereas 0.18 electrons are in $|m|$=1
orbitals. In the majority spin band major change was seen in $|m|$=1
state, whereas there was no meaningful modification found in other
two states like in unsupported structure. The m-resolved DOS of
Ti/NiAl(110) presents that the net magnetic moments originated from
$|m|$=2, $|m|$=1, and $|m|$=0 states are approximately 0.58, 0.36,
and 0.25 $\mu_B$, respectively. We observed that the charge transfer
due to the hybridization effect is mainly occurred in $d_{xz}$ and
$d_{yz}$ states and the $d_{3z^2-r^2}$ state is almost intact even
in the presence of substrate. Consequently, we conclude that the
$d_{x^2-y^2}$ and $d_{xy}$ states are most important in
ferromagnetism of 1D Ti atomic chain on NiAl(110), whereas $|m|$=0
and $|m|$=1 orbitals manifest almost the same contribution to the
ferromagnetism, but less important in ferromagnetism.

We now present the results of magnetic anisotropy of 1D Ti
nanowires. As well known, two physical origins can account for the
magnetic anisotropy of materials; magnetic dipolar interaction, so
called shape anisotropy and magnetocrystalline anisotropy (MCA) due
to spin-orbit interaction. It is straightforward to calculate the
shape anisotropy and the magnetic dipolar interaction always prefers
magnetization along the chain axis in 1D structure. The MCA arising
from the spin-orbit interaction requires very accurate method for
numerical calculations since the MCA is very sensitive to change in
the underlying electronic structure. In our calculations the
spin-orbit interaction is treated in second variational approach
based on the ground state properties obtained from semi-relativistic
calculations \cite{soc} and we used torque method \cite{torque} for
the calculation of magnetic anisotropy. Since the wire is placed
along the x-axis, we can write the expression of MAE as \eb
E=E_0-sin^2\theta(E_1+E_2cos^2\phi) \ee where $\theta$ is a polar
angle measured from the chain axis and $\phi$ is an azimuthal angle
measured from y-axis. Then, one can easily obtain the relations
$E_x-E_z=E_1$, $E_y-E_z=-E_2$, and $E_x-E_y=E_1+E_2$. We then
directly investigate $E_1$ and $E_2$ via torque method. It was found
that $E_1=13$ $\mu$eV and $E_2=-20$ $\mu$eV, so the direction of
magnetization due to the spin orbit interaction is perpendicular to
the chain axis (surface normal in our geometry). On the other hand
we found that the shape anisotropy due to magnetic dipolar
interaction is 33 $\mu$eV and the direction of magnetization is
along x-axis as mentioned. Overall, we conclude that the direction
of magnetization of 1D Ti nanowire is in x-z plane and its angle is
$21^o$ with respect to x-axis

In summary, we for the first time theoretically investigated the
existence of magnetic phase in 1D Ti atomic chain. It was found that
free standing Ti nanowire has magnetic moment of 1.88 $\mu_B$ and
very interestingly the magnetic moment of Ti atom was still
maintained even on the NiAl(110) surface with the magnitude of 1.24
$\mu_B$. We realized that the spin polarization of $d_{x^2-y^2}$ and
$d_{xy}$ states is the most essential in the observation of
ferromagnetic ground state. We also found that the magnetic dipolar
interaction is more important than the spin-orbit interaction in the
study of anisotropy. We hope our results stimulate further
experimental verification.

This work was supported by grant No. R01-2005-000-11001-0 from the
Basic Research Program of the Korea Science $\&$ Engineering
Foundation.

\label{111}

\newpage
\begin{figure}[t]
\caption{Schematic illustration of unit cell of Ti/NiAl(110)
considered in our calculations. The numbers indicate atomic type}
\end{figure}
\begin{figure}[t]
\caption{DOS of majority and minority spin electrons: (a) free
standing Ti (b) Ti/NiAl(110).}
\end{figure}
\begin{figure}[t]
\caption{m-resolved DOS of Ti: (a) free standing Ti (b)
Ti/NiAl(110). }
\end{figure}

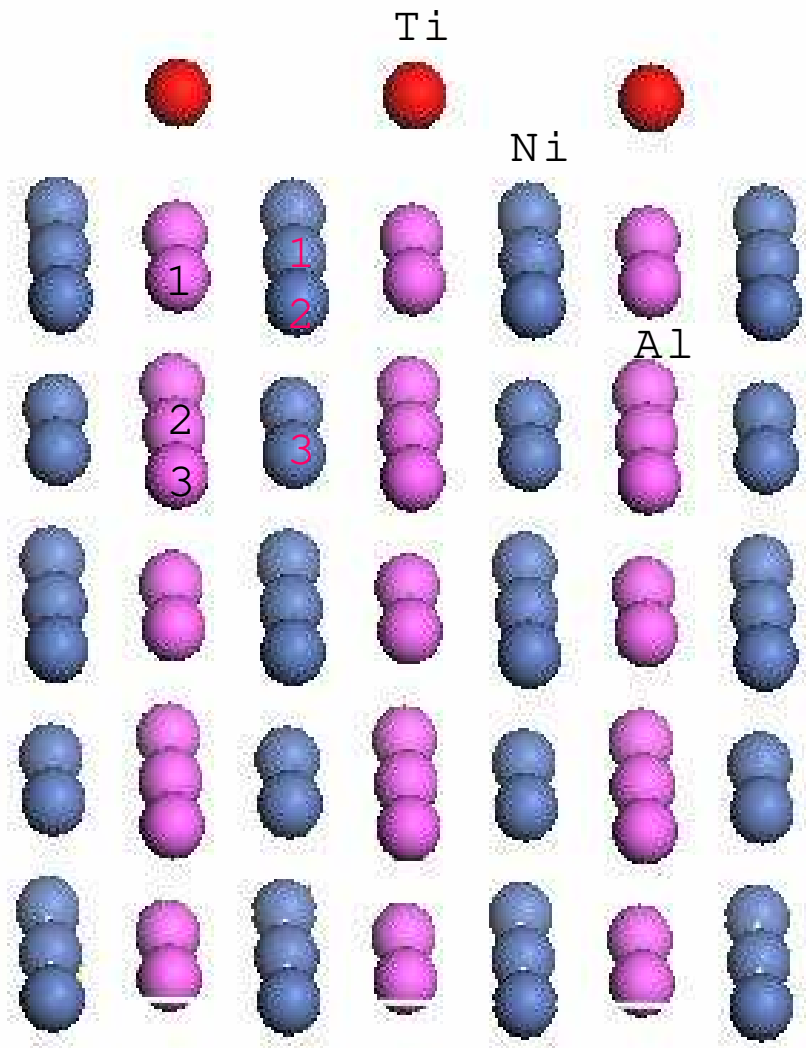
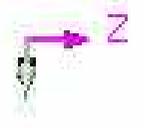
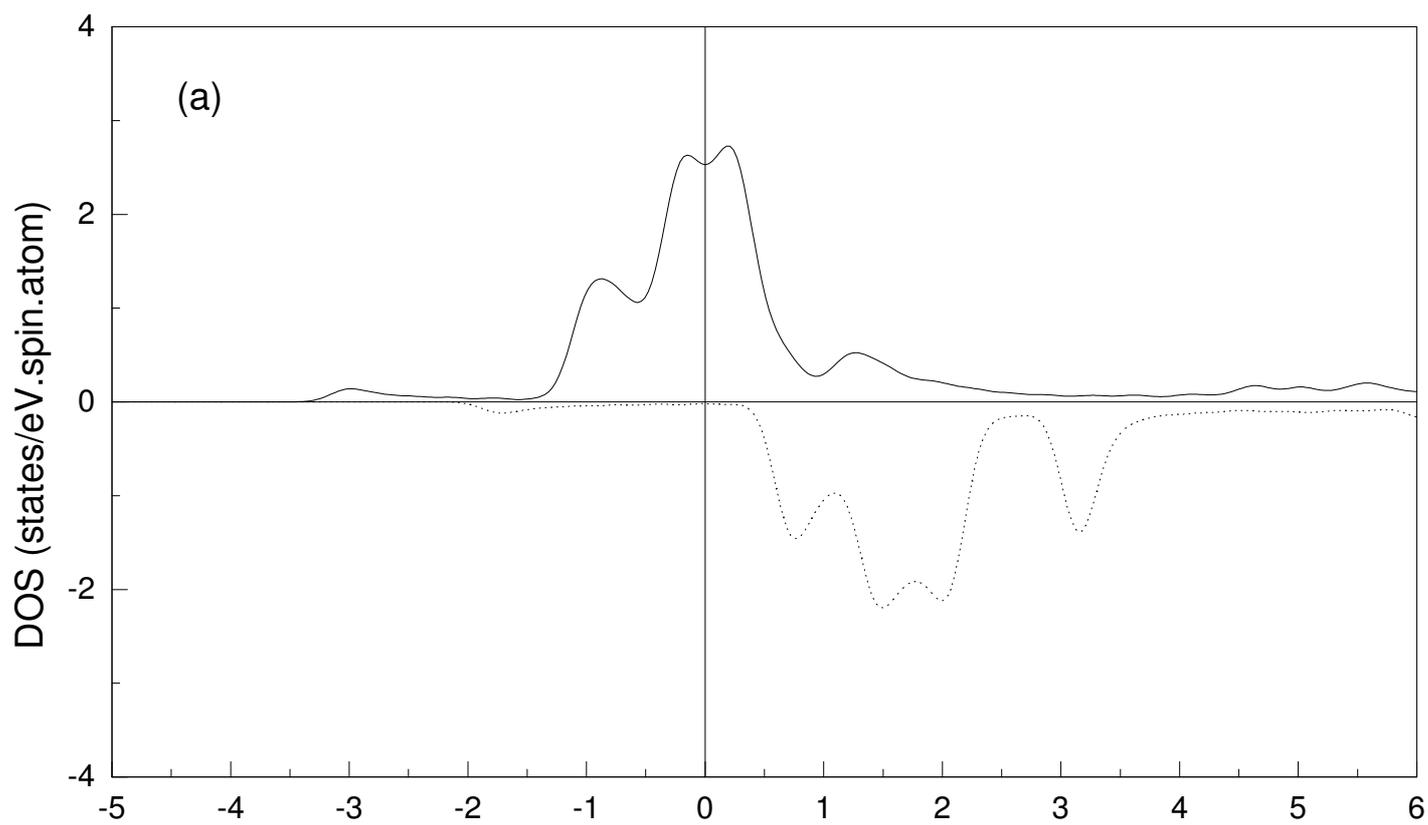
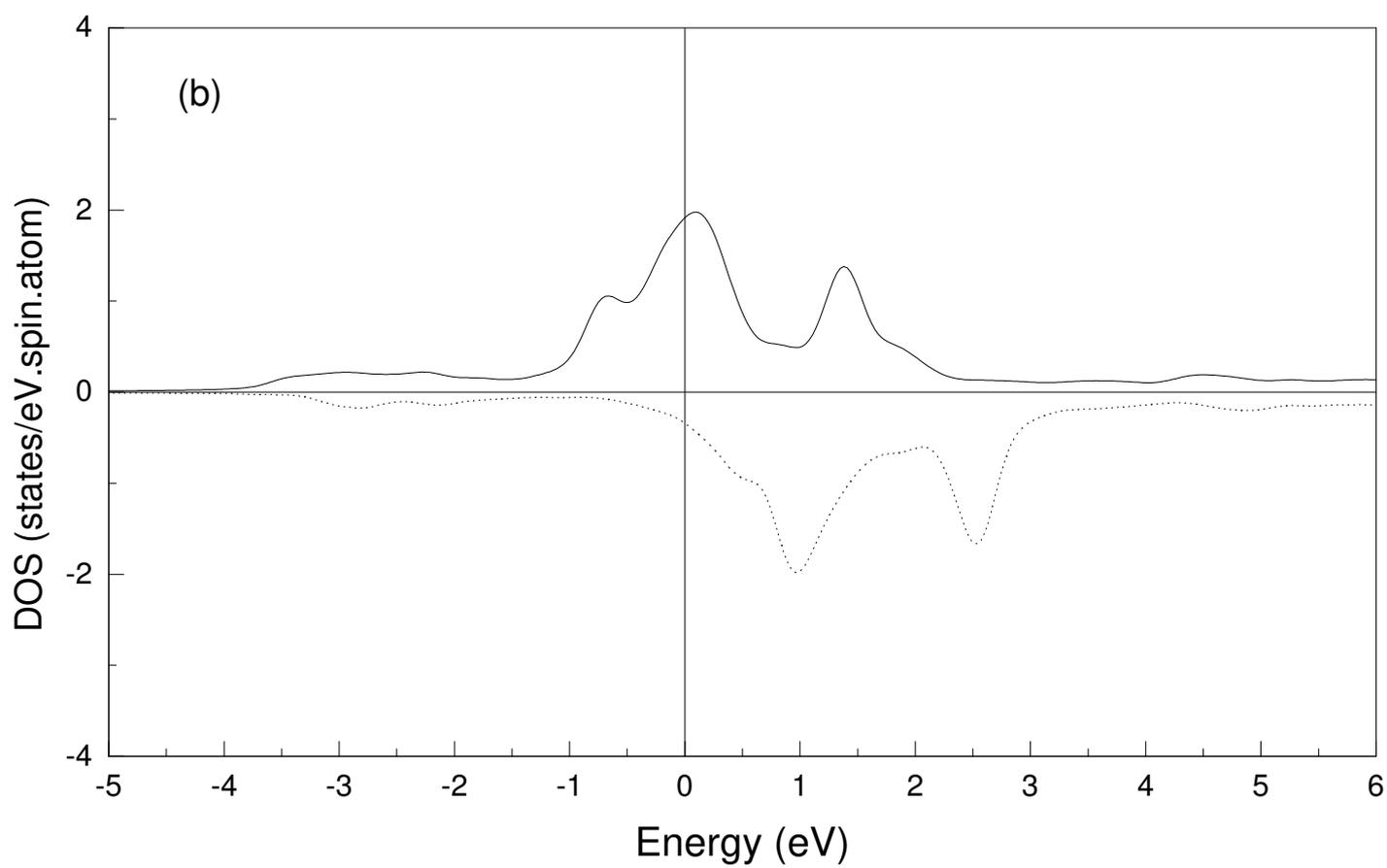
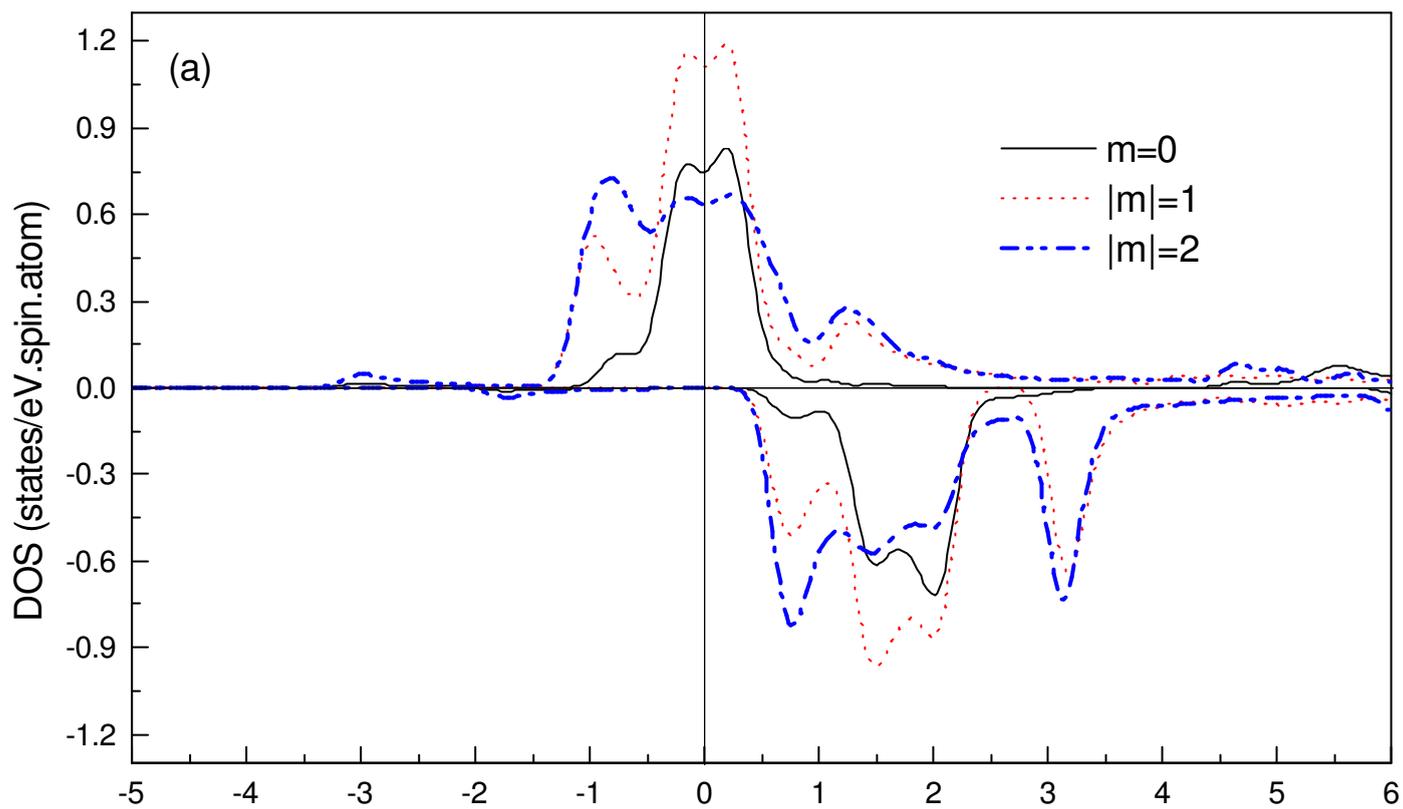
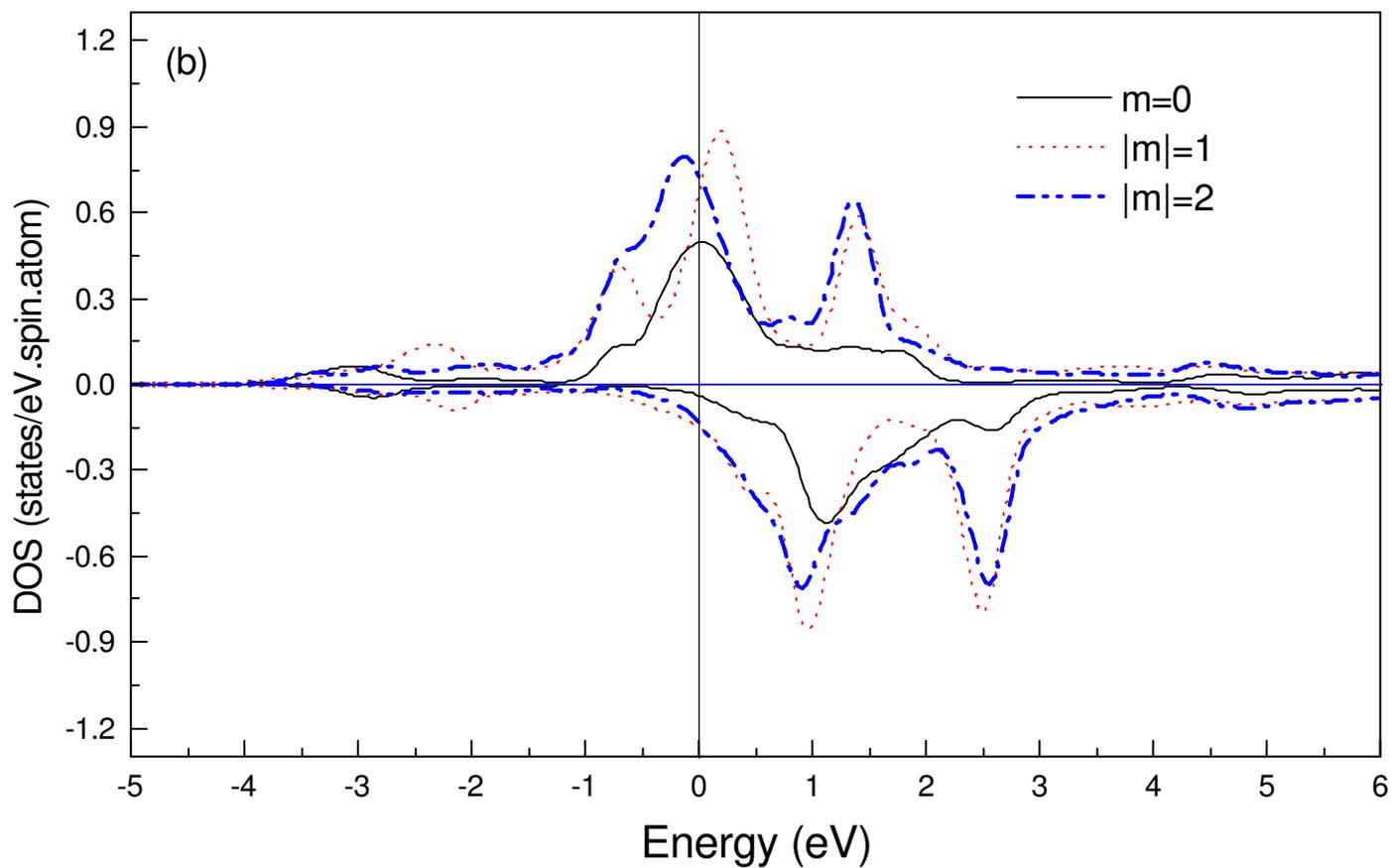
\end{document}